\RequirePackage{fix-cm}
\RequirePackage{amsmath}
\documentclass[smallextended]{svjour3}       % onecolumn (second format)

\usepackage{amssymb,amsfonts}
\usepackage{xcolor}
\smartqed
\usepackage{algorithmic}
\usepackage{graphicx}
\usepackage{textcomp}
\usepackage{xcolor}
\usepackage[disable]{todonotes}
\usepackage{fancybox,mdframed}
\usepackage{blindtext}
\usepackage{changepage} 
\usepackage{tikz}
\usetikzlibrary{arrows, shapes}
\usetikzlibrary{arrows.meta}
\usetikzlibrary{fit}
\usepackage{booktabs}
\usepackage{multirow}
\usepackage{lscape}
\usepackage{booktabs}
\usepackage{colortbl}
\usepackage[bookmarks=false]{hyperref}
\usepackage{graphicx}
\usepackage{balance}
\usepackage{lastpage}
\usepackage{float}
\usepackage{natbib}
\usepackage{afterpage}

\newcommand{\MyBox}[1]{\vspace{3mm}\noindent\framebox[\columnwidth][c]{\parbox[b]{0.95\columnwidth}{ #1 }}\vspace{3mm}}

\begin{document}

\title{Assessment of Off-the-Shelf SE-specific Sentiment Analysis Tools: An Extended Replication Study}
\titlerunning{Assessment of Off-the-Shelf SE-specific Sentiment Analysis Tools}

\author{
Nicole Novielli \and 
Fabio Calefato \and 
Filippo Lanubile \and 
Alexander Serebrenik
}


\institute{N. Novielli, F. Calefato, F. Lanubile \at
              University of Bari Aldo Moro, Italy \\
              \email{first.last@uniba.it}           %  \\
           \and
           A. Serebrenik \at
              Eindhoven University of Technology, The Netherlands\\
              \email{a.serebrenik@tue.nl}
}

\date{Received: date / Accepted: date}

\maketitle

\begin{abstract}

Sentiment analysis methods have become popular for investigating human communication, including discussions related to software projects. Since general-purpose sentiment analysis tools do not fit well with the information exchanged by software developers, new tools, specific for software engineering (SE), have been developed. 
We investigate to what extent off-the-shelf SE-specific tools for sentiment analysis mitigate the threats to conclusion validity of empirical studies in software engineering, highlighted by previous research. First, we replicate two studies addressing the role of sentiment in security discussions on GitHub and in question-writing on Stack Overflow. Then, we extend the previous studies by assessing to what extent the tools agree with each other and with the manual annotation on a gold standard of 600 documents. 
We find that different SE-specific sentiment analysis tools might lead to contradictory results at a fine-grain level, when used \textit{off-the-shelf}. Conversely, platform-specific tuning or retraining might be needed to take into account differences in platform conventions, jargon, or document lengths.

\end{abstract}


\section{Introduction}
\label{sec:introduction}

Sentiment analysis, i.e., the task of extracting the positive or negative semantic orientation a text~\citep{Pang:2008}, has emerged as a tool for empirical software engineering studies to mine emotions and opinions from textual developer-generated content~\citep{DBLP:journals/jss/NovielliBM19} in the `social programmer'~\citep{storey} ecosystem. Its popularity is also due to the availability of a plethora of sentiment analysis tools released for public use.

With a few notable exceptions~\citep{Blaz:2016,Ortu:2016,Panichella:2015}, early empirical studies in this field have exploited off-the-shelf general-purpose sentiment analysis tools.
%trained on non-software engineering documents.
However, general-purpose sentiment analysis tools tend to produce unreliable results in the  software engineering context as they are trained on documents from non-technical domains, such as film critics and product reviews.
Specifically, \citet{Jongeling:2017} tried to replicate previously published empirical studies and showed that the choice of the sentiment analysis tool has an impact on the validity of the conclusions.

To overcome such limitations, researchers have started developing SE-specific sentiment analysis tools to mine developers' emotions (e.g., ~\cite{Calefato:2018,Ahmed:2017,Islam:2017, Chen_Sentimoji:FSE2019}) and opinions (e.g., ~\cite{Lin2019,Uddin2017}). 
In previous benchmarking studies, ~\cite{Novielli:2018, Novielli:2020} showed how SE-specific customization gives a boost in accuracy in terms of both agreement with manual annotation and agreement among tools, 
provided that model-based annotation of emotions is implemented and that a robust gold-standard dataset for retraining is available.

In this paper, we go beyond the simple assessment of performance of SE-specific tools and their agreement, which is the main goal of the aforementioned benchmarking studies.
Here, we investigate to what extent the SE-specific retraining operated by the original authors of sentiment analysis tools enables researchers to address the threat to conclusion validity induced by the use of general-purpose tools~\citep{Jongeling:2017}. Specifically, we aim at understanding if we can reliably and safely reuse SE-specific tools \textit{off-the-shelf}, i.e., without further tuning or training beyond that performed by the original authors on technical texts. 

As such, we formulate our first research question as follows 
\begin{itemize}
	\item \textit{RQ1: Does the choice of the sentiment analysis tool introduce a threat to conclusion validity in a software engineering study?}
\end{itemize}

To address RQ1, we replicate two previously published studies on sentiment analysis in software engineering. We decided to perform exact, dependent replications \citep{Shull2008} to mitigate threats to validity that are inherent in replications. As such, we select two studies in which a subset of the authors of the current paper were involved as co-authors. By doing so, we are able to replicate the original study design, use the same dataset as the original studies, and change only the sentiment analysis tools used to mine the developers' sentiment.
Specifically, we replicate the analysis of sentiment in GitHub security discussion by~\cite{Pletea:2014} and the study of the role of emotions in effective question-writing in Stack Overflow by~\cite{Calefato:2018:IST}. First, we replicate each study with four SE-specific tools and compare the conclusions we obtain. Then, we compare the findings obtained with SE-specific tools with those previously published in the original studies to verify whether they still hold. 

Other than suggesting a threat to conclusion validity, \citeauthor{Jongeling:2017} showed that general-purpose sentiment analysis tools also disagree with human annotation and with each other. To get further insights in this direction, but this time focusing on SE-specific sentiment analysis tools, we inherit and revise two additional research questions from~\citet{Jongeling:2017}:

\begin{itemize}    
	\item \textit{RQ2: To what extent do results from different SE-specific sentiment analysis tools agree with each other?}
	\item \textit{RQ3: To what extent do different SE-specific sentiment analysis tools agree with human raters?} %(former RQ1)
\end{itemize}

We address RQ2 by measuring the agreement between the SE-specific tools themselves, based on the manual annotation of a subset of 600 documents randomly selected from the original datasets. We also use the obtained gold standard to address RQ3 by computing the agreement of SE-specific tools with the manual labels. We release the annotated corpus\footnote{https://figshare.com/s/faecffd83e65155ebc2a} as an additional contribution of the study to encourage further studies on emotion detection in software engineering. 

This paper enhances the state of the art on sentiment analysis in empirical studies in software engineering by furthering our understanding of \textit{off-the-shelf} tool reuse, even in presence of SE-specific tuning. We show how the choice of the specific sentiment analysis tools might lead to contradictory results, thus representing a threat to conclusion validity, when the analysis is at a fine-grained level. Our results suggest that SE-specific fine-tuning of sentiment analysis tools to the software engineering domain might not be enough to improve accuracy, and platform-specific tuning or retraining might be needed to adjust the model performance to the shifts in lexical semantics due to different platform jargon or conventions. Further fine-tuning should be implemented in order to adjust the sentiment operationalization in line with the specific research goals.

The remainder of the paper is organized as follows. In Section~\ref{sec:background}, we review the related work and describe the SE-specific sentiment analysis tools that we employ in our study. In Section~\ref{sec:original}, we summarize the two studies that we replicate, including their goals, settings, and results.
We report the results of our replications in Section~\ref{sec:replication} and the results of the analysis of the agreement between tools and with the manual labels in Sections~\ref{sec:agreement} and~\ref{sec:manual}, respectively. In Section~\ref{sec:discussion}, we discuss our findings while in Section~\ref{sec:threats} we assess the main limitations of the current study. Finally, we provide conclusions in Section~\ref{sec:conclusion}.

\section{Background}\label{sec:background}
\subsection{Related work}
\label{sec:related} 

In recent years, a trend has emerged to use sentiment analysis as a new tool for empirical studies in software engineering research. Other than the studies by \cite{Pletea:2014} and \cite{Calefato:2018:IST} that we replicate in this paper, several others have applied sentiment analysis to mine emotions of software developers from their communication traces. In particular, researchers investigated the role of affect in social software engineering by applying sentiment analysis to the content available in collaborative software development platforms such as GitHub~\citep{Guzman:2014,Sinha:2016} and Jira~\citep{Ortu:2016,Mantyla:2016,Murgia2018}. In the scope of requirements engineering research, sentiment analysis has been also leveraged to mine users' opinions about software products from their reviews in the app stores~\citep{Panichella:2015,KurtanovicM18}, from user-generated contents in microblogging platforms~\citep{GuzmanAS16} and customers' tickets~\citep{Werner2018}. 

In spite of the popularity of the topic, only a few papers report about replicated studies on sentiment analysis in software engineering. Our study builds upon previous research by \cite{Jongeling:2017} who compared the predictions of general-purpose, off-the-shelf sentiment analysis tools, showing that not only they disagree with human annotation, but also with each other. Given the disagreement among these tools, Jongeling and colleagues conducted a replication of previous studies on sentiment analysis in software engineering to understand to what extent the choice of an instrument affects the results. They observed  contradictory findings and concluded that previous studies' results cannot be replicated when different, general-purpose sentiment analysis tools are used, i.e., the instrument choice can induce threats to conclusion validity. 

\cite{Novielli:2018} investigated to what extent different SE-specific sentiment analysis tools agree with gold standard annotations of developers' emotions and with each other. To this aim, they performed a benchmarking study aimed at assessing the performance of three SE-specific sentiment analysis tools on four gold standard datasets including developers' communication traces from collaborative platforms, such as Stack Overflow and Jira. The performance of each tool was assessed on a held-out test set extracted from the gold standard, which was also used to build the baseline performance represented by a general-purpose tool. The findings of \citeauthor{Novielli:2018} indicate that reliable sentiment analysis in software engineering is possible, provided that manual annotation of gold standards is inspired by theoretical models of affect. Furthermore, they found that, regardless of the approach adopted for annotation, SE-specific customization/retraining does provide a boost in accuracy with respect to the baseline approach represented by  SentiStrength~\citep{Thelwall:2012}, an off-the-shelf general-purpose tool. Based on the findings of their qualitative error analysis, \citeauthor{Novielli:2018} suggest that we  should be aware that tools and datasets are built by having in mind specific research goals and different conceptualization of affect. Thus, a sanity check is always recommended to assess the suitability of existing tools with respect to the research goals. 

\cite{Mantyla:2017} replicated a previous study on the identification of burnout and productivity of software developers~\citep{Mantyla:2016}. The original study was performed by using a general-purpose lexicon of 14,000 English words with known emotion scores, including the arousal associated to each word.
In the replication study the authors built a Software Engineering Arousal lexicon (SEA) to correctly measure the arousal, i.e., the emotional activation (calm vs. excited), by means of linguistic analysis of developers' communication traces. 
They validated SEA and its ability to correctly capture the emotional arousal of developers from the dataset 700,000 Jira issue reports containing over 2,000,000 comments used in the original study~\cite{Mantyla:2016}. 
The results of the replicated study confirm previous findings that emotion-related metrics can be used to identify different types of issue reports as well as their priority. Furthermore, the replication with SEA shows that an SE-specific lexicon is able to better differentiate between issue priorities. In particular, a unified approach, combining SEA with the general  purpose  lexicon  used in the original study, offers  a clear  improvement  over  previous  work.

Overall, the results from previous studies suggest that SE-specific tools are required to overcome the limitations of general-purpose tools when applied to the software engineering domain. Furthermore, empirical evidence was provided that SE-specific tuning of tools improves the accuracy of sentiment analysis in empirical software engineering studies.

\subsection{SE-specific Sentiment Analysis Tools}
\label{sec:tools}

Sentiment analysis is the task of extracting positive, negative, or neutral orientation of opinions and emotions conveyed by a text. Despite the popularity of general-purpose  sentiment analysis tools, a consensus in the research community is that such tools are not appropriate for detecting  emotions in developers' technical discussions~\citep{Jongeling:2017, Novielli:2015, Lin:2018}. As such, researchers have implemented and released their own tools specifically customized for the software engineering domain. 
To enable comparison with previous findings, we select four tools that were already included in previous benchmarking studies on sentiment analysis in software engineering \citep{Novielli:2018, Novielli:2020}. 
Two of the tools are lexicon-based, namely SentiStrength-SE~\citep{Islam:2017} and DEVA~\citep{Islam:2018}, that is they rely on sentiment lexicons including polarity scores at the word level. The other two, namely Senti4SD~\citep{Calefato:2018} and SentiCR~\citep{Ahmed:2017}, implement a supervised approach and are trained on manually labeled gold standards.

\begin{table*}[b]
	\begin{center}
		\caption{Examples of the annotation produced by the SE-specific sentiment analysis tools.}
		\label{tab:examplesToolPredictions}
		\resizebox{\textwidth}{!}{
			\begin{tabular}{p{6cm}|c|cc|c|c}
				& \textbf{Senti4SD}        & \multicolumn{2}{c}{\textbf{SentiStrength-SE}}     & \textbf{DEVA}                & \textbf{SentiCR}     \\
				\hline
				\textbf{Example documents from our datasets} & \textit{Polarity label} & \textit{Pos Score} & \textit{Neg Score}  & \textit{Score} & \textit{Label} \\
				\hline
				I have Ubuntu 14.04 , I've installed JULIA, and then I tried to install IJULIA, writting Pkg.add("IJulia") on julia, but this appears: Do any one knows what to do? What I want to achieve is to run ipython notebook --profile julia                                       & neutral                              & 1                  & -1                & 1                           & 0                       \\
				\hline
				I'm trying to use this cod for checking expiry date. I'm using Code before to open different file and copy cells as below. Please can any one help why I have Run time Error 13 Type mismatch. Any ideas ?? I was trying to put If empty cell exit sub but still that same :( & negative                            & 1                  & -2                 & -2                           & -1                         \\
				\hline
				Here is my code: \} It works fine so far without warning or errors. The only problem is, in the applications it shows for example Dauer: 0:4:3 instead of 00:04:03. Does anyone know how I can fix this? Thank you very much... :)                                          & positive                           & 3                  & -1                    & 3                         & 0           \\
				\hline
			\end{tabular}
		}
	\end{center}
\end{table*}

\textit{SentiStrength-SE} \citep{Islam:2017} is an SE-specific adaptation of the general-purpose tool SentiStrength~\citep{Thelwall:2012}. 
It leverages a manually adjusted version of the SentiStrength lexicon and implements \textit{ad hoc} heuristics to correct the misclassifications observed when running it on the dataset by~\citet{Ortu:2016}. The sentiment scores of words in the lexicon were manually adjusted to reflect the semantics and neutral polarity of domain words such as 'support' or 'default'. SentiStrength-SE assigns an integer value between $1$ and $5$ for the positivity of a text, $p$ and, analogously, a value between $-1$ and $-5$ for the negativity $n$ (respectively, the positive and negative scores in second and third columns of Table~\ref{tab:examplesToolPredictions}). The evaluation performed by the authors shows that SentiStrength-SE achieves .74 precision and .85 recall, which are higher than the performance of the general purpose tool SentiStrength on technical texts. 

\textit{DEVA} \citep{Islam:2018} leverages a lexicon-based approach for the identification of both emotion activation (arousal) and polarity from text. To this end, the tool uses two separate dictionaries developed by exploiting a general-purpose lexicon as well as one specific for software engineering text. To further increase its accuracy, DEVA also includes several heuristics, some of which are borrowed from SentiStrength-SE. Given an input text, the tool issues an integer score defined in the interval $\pm[1, 3]$, where values $<-1$ are considered negative and those $>1$ are considered positive; otherwise values equal to $\pm1$ are neutral (see fourth column of Table~\ref{tab:examplesToolPredictions}). For the empirical evaluation, the authors constructed a ground-truth dataset consisting of 1,795 JIRA issue comments, manually annotated by three human raters, on which DEVA was found to achieve a precision of .82 and a recall of .79.  

\textit{Senti4SD} \citep{Calefato:2018} is a polarity classifier trained to support sentiment analysis in developers' communication channels using supervised machine learning. It leverages a suite of features based on Bag of Words (BoW), sentiment lexicons, and semantic features based on word embedding.  Along with the toolkit, the authors distribute a classification model, trained and validated on a gold standard of about 4K questions, answers, and comments from Stack Overflow, and manually annotated for sentiment polarity. Furthermore, the toolkit provides a training method that enables the customization of the classifier using a gold standard as input. Given any text in input, the tool returns a predicted label in $\{\mbox{\it positive}, \mbox{\it neutral}, \mbox{\it negative}\}$ (see the first column of Table~\ref{tab:examplesToolPredictions}). Compared to the performance obtained by SentiStrength on the same Stack Overflow test set, Senti4SD reduces the misclassifications of neutral and positive posts as emotionally negative (precision = .87, recall = .87, F1=.87). The authors also report a good performance (F1=.84) of the tool with a minimal set of training documents.

\textit{SentiCR} \citep{Ahmed:2017} is a supervised tool that leverages a feature vector generated by computing tf-idf for Bag-of-Words (BoW) extracted from the input text. SentiCR implements basic preprocessing of the raw input text to expand contractions, handle negations and emoticons, remove stop-words, derive word stems, and remove code snippets. Furthermore, it performs SMOTE~\citep{Chawla:2002} to handle the class imbalance in the train set. Given a text in input, SentiCR issues either $-1$, for sentences predicted as negative, or $0$ for non-negative ones---i.e., as opposed to Senti4SD, SentiStrength-SE and DEVA, SentiCR does not distinguish between positive and neutral content (see the last column of Table~\ref{tab:examplesToolPredictions}). ~\cite{Ahmed:2017} report a mean accuracy of .83 for the tool. The currently distributed version implements a training approach based on Gradient Boosting Tree and requires a training set as an input in order to retrain the model and use it on the target document collection. 

In our replications, we follow the methodology of \citet{Jongeling:2017} and use all the SE-specific tools as `off-the shelf' resources.
For the two supervised tools Senti4SD and SentiCR we use the classifier trained on the original gold standard by their respective authors.  
Analogously, we use Sentistrength-SE and DEVA as they are originally implemented, that is leveraging the lexical resources as they are distributed by the authors after fine-tuning for the software engineering domain.

\section{Original Studies}
\label{sec:original}

In this section, we summarize the two original studies replicated in this paper, giving an overview of their research goals, settings, methodology, and results, following the  replication report guidelines for software engineering studies by~\cite{Car10}. We aim at performing exact replications ~\citep{Shull2008} of the studies, except for the use of a SE-specific sentiment analysis tool. Specifically, the original study of~\cite{Pletea:2014} used NLTK\footnote{Natural Language Processing APIs and Python NLTK Demos, \url{ http://text-processing.com}}, while the original study by~\cite{Calefato:2018:IST} used SentiStrength.\footnote{http://sentistrength.wlv.ac.uk} In our replications, we employ the four SE-specific tools reviewed in Section~\ref{sec:tools}. 

\subsection{Sentiment Analysis of Security Discussions on GitHub (\citeauthor{Pletea:2014})}
\label{sec:original1}
\textit{Research goals}. The first study we have chosen to replicate is the one by~\cite{Pletea:2014}, which analyzes discussions about security in GitHub. The authors conclude that security-related threads contain more negative emotions than non-security related ones. 

\textit{Dataset and methodology}.
The dataset used by~\citeauthor{Pletea:2014} consists in 60,658 commits and 54,892 pull requests from the MSR 2014 Mining Challenge Dataset~\citep{Gousios:2013}, consisting of data for 90 GitHub projects (repositories) and their forks.  

In their analysis, \citeauthor{Pletea:2014} distinguish between \textit{i}) commits comments, \textit{ii}) commit discussions, \textit{iii}) pull request comments, and \textit{iv}) pull request discussions. 
The term `discussion' refers to the entire collection of comments belonging to a commit  or a pull request.
Furthermore, the authors distinguish between  \textit{a}) security-related and \textit{b}) non-security related comments and discussions, thus resulting in eight different categories of texts. In this study, we use the same dataset used in the original study, as reported in Table~\ref{tab:datasetPletea}.

\begin{table}[htb]
	\begin{center}
		\caption{Dataset of security-related comments and discussions}
		\label{tab:datasetPletea}
		\begin{tabular}{llll}
			\textbf{Data source} & \textbf{Topic} & \multicolumn{2}{c}{\textbf{Distribution}}\\
			\hline
			& & \textit{Comments} & \textit{Discussions}\\
			\hline 
			Commits & Security      & 2509 (4.13\%) & 1706 (9.28\%)  \\
			& Total         & 60658         & 18378          \\
			\hline
			Pull Requests & Security      & 1801 (3.28\%) & 1091 (11.36\%) \\
			& Total         & 54892         & 9601          \\
			\hline
		\end{tabular}
	\end{center}
\end{table}

To address the research questions, the authors apply NLTK to GitHub comments and discussions.  Given  an  input text, NLTK outputs the probabilities that the text is neutral, negative, or positive as well as a polarity label summarizing the three scores. The probabilities for positive and negative sum up to 1 while the probability for neutral is computed \textit{per se}. A text is classified as neutral if the neutral probability is greater than chance. Conversely, it is either classified as positive or negative based on the highest probability score.
Depending on the aggregation considered (comments vs. discussions),~\citeauthor{Pletea:2014} provide in input to NLTK either the individual comments or the entire pull request/commit discussions. Hence, the polarity level is issued at the level of either individual comment or %at the level 
the entire discussion, according to the chosen unit of analysis. 

\textit{Findings}. The authors report that negative sentiment occurs more frequently in security related texts for both commit/pull request comments and discussions. Moreover, they observe that the NLTK results are characterized by both strong negative and strong positive sentiment. As such, \citeauthor{Pletea:2014} conclude that the security-related discussions tend to be more emotional than those not related to security.

A first replication of this study was performed by~\cite{Jongeling:2017} using both NLTK and SentiStrength.
SentiStrength~\citep{Thelwall:2012} is a general-purpose sentiment analysis tool widely adopted in social computing at the time of their study.
SentiStrength is capable of dealing with short informal text, including abbreviations, intensifiers, and emoticons. Based on the assumption that a sentence can convey mixed sentiment, SentiStrength outputs both positive and negative sentiment scores for an input text. Positive sentiment scores range from +1 (absence of positive sentiment) to +5 (extremely positive) while negative sentiment scores range from -1 (absence of negative sentiment) to -5 (extremely negative).

Comparing the results obtained using NLTK in the original study by~\citeauthor{Pletea:2014} and SentiStrength, \citeauthor{Jongeling:2017} observe differences in term of polarity label distributions. Overall, NLTK issues more negative labels, thus suggesting that both comments and discussions are predominantly negative. Conversely, SentiStrength classifies the texts as predominantly neutral. In spite of these differences, overall the study of \citeauthor{Jongeling:2017}  confirms the findings originally reported by~\citeauthor{Pletea:2014}, i.e., security pull request/commit comments and discussions convey more negative sentiment than those revolving around other topics.

\subsection{How to Ask for Technical Help (\citeauthor{Calefato:2018:IST})}
\label{sec:original2}

\textit{Research goals}. The second study we have chosen to replicate is the one by~\cite{Calefato:2018:IST}, which investigates how information seekers can increase the chance of eliciting a successful answer to their questions on Stack Overflow---i.e., the best answer marked as the accepted solution by the original asker. To achieve their research goal, they develop a framework of factors influencing the success of questions in Stack Overflow. The framework is built upon the community recommendations and the results from previous research in this field. Specifically, the authors focus on actionable factors that can be acted upon by software developers when writing a question to ask for technical help, namely \textit{affect} (i.e., the positive or negative sentiment conveyed by text), \textit{presentation quality}, and \textit{time}. The asker's \textit{reputation} is also included in the framework as a control factor. 
As far as sentiment is concerned, the authors investigate the correlation between expressing sentiments in a question on Stack Overflow and the question probability of eliciting a successful answer. 

\textit{Dataset and methodology}. \citeauthor{Calefato:2018:IST} analyze a dataset of 87,373 questions extracted from the official Stack Overflow dump, using logistic regression to estimate the probability of a successful question. 
The authors operationalize the framework by defining a set of metrics that are used as predictors in a logistic regression analysis. As for the affect factor, they operationalize it using two metrics, namely \textit{Positive Sentiment} and \textit{Negative Sentiment}, whose combination of values represents the overall sentiment orientation of a question. Specifically, they measure the overall positive polarity (\textit{Positive Sentiment}) and negative polarity (\textit{Negative Sentiment}) of the question body. 

To capture the sentiment of questions, \citeauthor{Calefato:2018:IST}  use SentiStrength ~\citep{Thelwall:2012}. 
They discretize the sentiment scores by treating Positive Sentiment and Negative Sentiment as Boolean variables. 
Specifically, they encode whether a question shows in its lexicon a positive (positive score in [+2, +5]) or negative (negative score in [-2, -5]) affective load. A question is neutral when positive sentiment score is +1 and negative sentiment score is -1.

\textit{Findings}. \citeauthor{Calefato:2018:IST} report that, regardless of user \textit{reputation}, successful questions are short, contain code snippets, and do not abuse with uppercase characters. As regards \textit{affect}, successful questions adopt a neutral emotional style. Based on that, they recommend avoiding both rudeness as well as positive sentiment in question writing. As regards \textit{presentation quality}, they recommend including example code snippets, writing concise questions, and using capital letters only where appropriate. Finally, they observe the impact of \textit{time} on the question success and advice Stack Overflow users to be aware of low-efficiency hours of the community.

\section{Replications with SE-specific sentiment analysis tools}
\label{sec:replication}

To address RQ1 (\textit{Does the choice of the sentiment analysis tool introduce a threat to conclusion validity in a software engineering study?}), 
we replicate the original studies by~\cite{Pletea:2014} and~\cite{Calefato:2018:IST} using the four SE-specific sentiment analysis tools reviewed in Section~\ref{sec:tools}. 
The mapping of the tool output to consistent polarity labels is described in Section~\ref{sec:toolMapping}. The results of the two replications are reported in Sections~\ref{sec:resultsPletea} and \ref{sec:resultsCalefato}, respectively.

\subsection{Mapping the Output of Tools to Polarity Labels}
\label{sec:toolMapping}
In our replications, we use all the SE-specific tools as `off-the shelf' resources. 
Since the tools issue heterogeneous outputs, we need to map them into polarity labels by enforcing the same operationalization of sentiment adopted in the original studies. 

In their study, \citeauthor{Pletea:2014} model sentiment using a single variable with three polarity classes, namely \textit{negative}, \textit{positive}, and \textit{neutral} (see Section~\ref{sec:original1}). The output of Senti4SD is consistent with such schema, so we do not need to apply any changes. Similarly, the scores produced by DEVA can be directly mapped into the corresponding negative ($\mbox{\it score} < 1$), positive ($\mbox{\it score} > 1$), and neutral ($\mbox{\it score} = 1$) labels. 
Because the output of SentiStrength-SE has the same structure as the output of SentiStrength, we adopt the  mapping implemented for SentiStrength in the former replication by~\cite{Jongeling:2017}: a text is considered positive when $p + n > 0$, negative when $p + n < 0$, and neutral if $p = -n$ and $p < 4$. Texts with a score of $p = -n$ and $p \geq 4$ are considered having an undetermined sentiment and, therefore, removed from the datasets. Finally, we translate the categorical scores of SentiCR into negative ($\mbox{\it score} = -1$) and non-negative ($\mbox{\it score} = 0$). 

\citeauthor{Calefato:2018:IST} use two Boolean values (\textit{Positive Sentiment} and \textit{Negative Sentiment}) to indicate the presence of positive and negative emotions, respectively. \textit{Neutral} is modeled by assigning `false' to both values. To represent mixed cases both values are equals to `true' (see Section~\ref{sec:original2}). As for Senti4SD and DEVA, we implemented a direct mapping by setting 'true'/'false' values based on the output of the tool. As for SentiStrength-SE, positive and negative sentiment are set to `true' if $\mbox{\it positive score} > 1 $ and $\mbox{\it negative score} < -1 $, respectively, in line with the original study. Finally, the categorical scores of SentiCR are used to indicate the presence of negative sentiment. Conversely, the positive sentiment cannot be modeled in the empirical setting using SentiCR as this tool does not distinguish between positive and neutral.

\subsection{Replication of Pletea et al.}
\label{sec:resultsPletea}
As in the original study, we compute the proportions of negative, neutral, and positive in the \textit{security} comments and discussions for both commits and pull requests. The resulting values are compared with the proportions of negative, neutral, and positive sentiment in the comments and discussions revolving around other topics, i.e. the \textit{rest} of the texts in the dataset.
We report the results from the replication and the comparison with the original work by Pletea et al. in Tables~\ref{tab:resultsCommit} and~\ref{tab:resultsPullRequest} for commits and pull requests, respectively. For the sake of completeness, we also report the results of the former replication of the work of Pletea et al. performed by ~\cite{Jongeling:2017}, which used both NLTK and SentiStrength.

\begin{table}[tb]
	\caption{Commit sentiment analysis statistics. The largest polarity group per study and topic is typeset in boldface. We report in \textit{Italic} the higher percentage of negative comments between security and rest of the topics per study.}
	\label{tab:resultsCommit}
	\begin{center}
		\begin{tabular}{llrrr}
			\textbf{Tool}            & \textbf{Topic}     & \textbf{Negative} & \textbf{Neutral} & \textbf{Positive} \\
			\hline
			\multicolumn{5}{l}{\textbf{Commit Discussion}}\\
			\hline
			\multicolumn{5}{c}{\textit{Original study by~\cite{Pletea:2014}}}\\
			NLTK    & Security  & \textbf{\textit{72.52\%}}  & 10.88\% & 16.58\%  \\
			& Rest     & \textbf{54.28\%}  & 20.37\% & 25.33\%  \\
			\multicolumn{5}{c}{\textit{Replication by ~\cite{Jongeling:2017}}}\\
			NLTK       & Security & \textbf{\textit{70.16\%  }}& 12.79\% & 17.05\%  \\
			& Rest     & \textbf{52.89\% } & 21.50\% & 25.61\%  \\
			SentiStrength      & Security & \textit{30.66\%}  & \textbf{42.92\%} & 26.40\%  \\
			& Rest     & 24.13\%  & \textbf{43.92\%} & 31.94\%  \\
			\hline
			\multicolumn{5}{c}{\textit{Our replication with SE-specific tools}}\\
			SentiStrength-SE       & Security & \textit{24.73\%}  & \textbf{43.96\%} & 31.30\%  \\
			& Rest     & 17.07\%  & \textbf{51.61\%} & 31.31\%  \\
			DEVA          & Security & \textbf{\textit{40.80\%}}  & 19.69\% & 39.51\%  \\
			& Rest     & 31.79\%  & 32.20\% & \textbf{36.01\%}  \\
			SentiCR       & Security & \textbf{\textit{61.08\%}}  & \multicolumn{2}{c}{38.92\% (non-neg.)}         \\
			& Rest     & 37.07\%  & \multicolumn{2}{c}{\textbf{62.92\%} (non-neg.)}          \\
			Senti4SD          & Security & \textbf{\textit{36.93\%}}  & 27.72\% & 35.34\%  \\
			& Rest     & 22.23\%  & \textbf{43.07\%} & 34.69\%  \\
			\hline
			\multicolumn{5}{l}{\textbf{Commit Comments}}\\
			\hline
			\multicolumn{5}{c}{\textit{Original study by~\cite{Pletea:2014}}}\\
			NLTK    & Security & \textbf{\textit{55.59\%}}  & 23.42\% & 20.97\%  \\
			& Rest     & \textbf{46.94\%}  & 26.58\% & 26.47\%  \\
			\multicolumn{5}{c}{\textit{Replication by~ \cite{Jongeling:2017}}}\\
			NLTK        & Security & \textbf{\textit{55.96\%}}  & 22.88\% & 21.16\%  \\
			& Rest     & \textbf{46.89\%}  & 26.61\% & 26.50\%  \\
			SentiStrength        & Security & \textit{32.60\%}  & \textbf{46.95\%} & 20.44\%  \\
			& Rest     & 22.30\%  & \textbf{50.74\%} & 26.95\%  \\
			\hline
			\multicolumn{5}{c}{\textit{Our replication with SE-specific tools}}\\
			SentiStrength-SE & Security & \textit{15.46\%}  & \textbf{68.84\%} & 15.70\%  \\
			& Rest     & 12.27\%  & \textbf{66.85\%} & 20.88\%  \\
			DEVA          & Security & \textit{25.87\%} & \textbf{51.13\%} & 23.00\%  \\
			& Rest     & 21.10\%  & \textbf{51.38\%} & 27.52\%  \\
			SentiCR          & Security & \textit{28.93\%}  &\multicolumn{2}{c}{\textbf{71.06\%} (non-neg.)}\\
			& Rest     & 19.49\%  &\multicolumn{2}{c}{\textbf{80.50\%} (non-neg.)}       \\
			Senti4SD          & Security & \textit{17.58\%}  & \textbf{60.66\%} & 21.76\%  \\
			& Rest     & 11.20\%  & \textbf{63.47\%} & 25.33\\
			\hline
		\end{tabular}
	\end{center}
\end{table}

\begin{table}[tb]
	\begin{center}
		\caption{Pull request sentiment analysis statistics. The largest polarity group per study and topic is typeset in boldface. We report in \textit{Italic} the higher percentage of negative comments between security and rest of the topics per study.}
		\label{tab:resultsPullRequest}
		\begin{tabular}{llrrr}
			\textbf{Tool}                                   & \textbf{Topic}     & \textbf{Negative} & \textbf{Neutral} & \textbf{Positive} \\
			\hline
			\multicolumn{5}{l}{\textbf{Pull Request Discussion}}\\
			\hline
			\multicolumn{5}{c}{\textit{Original study by~\cite{Pletea:2014}}}\\
			NLTK                                    & Security & \textbf{\textit{81.00\%}}  & 5.52\%  & 13.47\%  \\
			& Rest     & \textbf{69.58\% } & 11.98\% & 18.42\%  \\
			\multicolumn{5}{c}{\textit{Replication by~\cite{Jongeling:2017}}}\\
			NLTK                                    & Security & \textbf{\textit{77.61\%}}  & 7.03\%  & 15.36\%  \\
			& Rest     & \textbf{67.43\%}  & 13.82\% & 18.76\%  \\
			SentiStrength                           & Security & \textit{30.80\%}  & \textbf{45.51\%} & 23.68\%  \\
			& Rest     & 24.15\%  & \textbf{51.17\%} & 24.67\%  \\
			\hline
			\multicolumn{5}{c}{\textit{Our replication with SE-specific tools}}\\
			SentiStrength-SE                        & Security & \textit{27.33\%}  & \textbf{46.63\%} & 26.04\%  \\
			& Rest     & 18.81\%  & \textbf{58.90\% }& 22.29\%  \\
			DEVA                                    & Security & \textbf{\textit{47.48\%}}  & 21.17\% & 31.35\%  \\
			& Rest     & 30.35\%  & \textbf{41.66\%} & 27.99\%  \\
			SentiCR                                 & Security & \textbf{\textit{76.08\%}}  & \multicolumn{2}{c}{23.92\% (non-neg.)}          \\
			& Rest     & \textbf{52.06\% } & \multicolumn{2}{c}{47.94\% (non-neg.)}        \\
			Senti4SD                                & Security & \textbf{\textit{49.31\%}}  & 31.90\% & 18.79\%  \\
			& Rest     & 28.26\%  & \textbf{49.55\% }& 22.18\%  \\
			\hline
			\multicolumn{5}{l}{\textbf{Pull Request Comments}}\\
			\hline
			\multicolumn{5}{c}{\textit{Original study by~\cite{Pletea:2014}}}\\
			NLTK                                    & Security & \textbf{\textit{59.83\%}}  & 19.09\% & 21.06\%  \\
			& Rest     & \textbf{50.16\%}  & 26.12\% & 23.70\%  \\
			\multicolumn{5}{c}{\textit{Replication by~ \cite{Jongeling:2017}}}\\
			NLTK                                    & Security & \textbf{\textit{59.67\%}}  & 18.83\% & 21.50\%  \\
			& Rest     & \textbf{49.81\%}  & 26.45\% & 23.74\%  \\
			SentiStrength                           & Security & \textit{25.66\%}  & \textbf{51.22\%} & 23.11\%  \\
			& Rest     & 18.14\%  & \textbf{62.87\% }& 18.97\%  \\
			\hline 
			\multicolumn{5}{c}{\textit{Our replication with SE-specific tools}}\\
			SentiStrength-SE                        & Security & \textit{11.60\%}  & \textbf{78.29\%} & 10.10\%  \\
			& Rest     & 8.58\%   & \textbf{79.97\%} & 11.45\%  \\
			DEVA                                    & Security & \textit{18.77\%}  & \textbf{63.91\%} & 17.32\%  \\
			& Rest     & 12.77\%  & \textbf{70.55\%} & 16.67\%  \\
			SentiCR                                 & Security & \textit{28.76\%}  & \multicolumn{2}{c}{\textbf{71.24\%} (non-neg.)}          \\
			& Rest     & 19.62\%  & \multicolumn{2}{c}{\textbf{80.37\%} (non-neg.}          \\
			Senti4SD                                & Security & \textit{11.60\%}  & \textbf{74.90\%} & 13.49\%  \\
			& Rest     & 7.28\%   & \textbf{77.70\%} & 15.01\% \\
			\hline                                        
		\end{tabular}
	\end{center}
\end{table}

From Tables~\ref{tab:resultsCommit} and~\ref{tab:resultsPullRequest}, we  notice that there is always a  larger proportion of negative polarity for security topics, regardless of the tool and the type of text. These results confirm the findings of the original study, as well as the findings of the replication by Jongeling and colleagues. 
In particular, we observe that different proportions of positive, negative, and neutral labels are issued by the different tools. Despite such differences, the original conclusion of \citeauthor{Pletea:2014} still holds: whether we consider comments or discussions, commits or pull requests, the proportion of negative sentiment among security-related texts is higher than among non-security related texts. 
As such, we can claim that, in the case of the replication of \citeauthor{Pletea:2014}, different SE-specific tools do not lead to contradictory conclusions for this study. %(RQ1.1). 
Furthermore, the choice of the sentiment analysis tool does not affect the conclusion validity of the results previously published. %(RQ1.2).

However, the differences in the proportions of positive, negative, and neutral labels issued by the various tools indicate the need for further reflection on the possible threats to conclusion validity that might be due to the 'off-the-shelf' use of sentiment analysis tools, beyond the confirmation of the high level findings of the original study. In both Table~\ref{tab:resultsCommit} and Table~\ref{tab:resultsPullRequest}, we observe such differences. 
First of all, we observe that SE-specific tools for sentiment analysis all tend to classify pull request and commit comments as predominantly neutral. This is in line with previous evidence~\citep{Calefato:2018} showing how the SE-specific tools are able to solve the negative bias of general purpose ones, due to the inability of the latter in dealing with technical jargon and domain specific semantics of words, such as `kill' or `save,' which are considered non-neutral outside the technical domain. As such, we do not observe the prevalence of negative comments reported in the original study of Pletea and colleagues, which used the general purpose tool NLTK. Even if SE-specific tools agree on classifying the comments as mainly neutral, we still observe differences in the percentages. This pertains to the problem of assessing the agreement between the SE-specific tools (RQ3), which we address in Section~\ref{sec:agreement}. 

As for pull request and commit discussions, overall we observe a lower  proportion of neutral labels, even when the same tool is adopted. For example (see the lower half of Table~\ref{tab:resultsCommit}), SentiStrength-SE classifies $68.84\%$ of security commit comments as neutral, $15.46\%$ as negative and $15.70\%$ as positive. The situation changes when discussions are analyzed as a whole, i.e., as a group of comments belonging to the same thread originated by the commit (see the upper part of Table~\ref{tab:resultsCommit}). In fact, SentiStrength-SE classifies $43.96\%$ of discussions as neutral, with a resulting higher percentage of negative ($24.73\%$) and positive ($31.30\%$). Same considerations hold for the other tools ad for the pull request analysis (Table~\ref{tab:resultsPullRequest}). This evidence suggests that different findings can be derived also depending on the level of granularity of the unit of analysis (comments vs. discussions, in this case).  

\subsection{Replication of Calefato et al.}
\label{sec:resultsCalefato}

In our replication, we use the dataset of the original study.\footnote{The original dataset can be downloaded from https://goo.gl/whZEWA}  As in the original study, we apply a logistic regression for estimating the correlation of each factors on the probability of success of a Stack Overflow question. In particular, we use the output of the four SE-specific sentiment analysis tools to recompute the metrics associated with the \textit{affect} and other factors of their framework, i.e., the positive and negative sentiment scores. 
Conversely, we do not recompute the metrics associated to the remaining factors and use those originally extracted in the original study, as they are distributed in the original dataset. Since SentiCR only distinguishes between negative and non-negative sentiment, in the replication using this tool we could only include the \textit{Positive Sentiment} among the predictors. 
In line with the original study, we treat the success of a question---i.e., the presence of an accepted answer---as the dependent variable and the metrics that operationalize each factors as independent variables. 

%\afterpage{\input{tables/ISTreplications.tex}}

\afterpage{\begin{landscape}
		\begin{table*}[tbh]
			\centering
			\begin{tabular}{l|p{3cm}|rr|rr|rr|rr|rr}
				\multirow{ 3}{*}{\textbf{Factor}}& \multicolumn{1}{c}{\multirow{3}{*}{\textbf{Predictor}}} & \multicolumn{10}{c}{\textbf{\textbf{Experimental setting: sentiment analysis tool used}}} \\
				&\multicolumn{1}{c}{} & \multicolumn{2}{c}{\begin{tabular}[c]{@{}l@{}}SentiStrength\\ \cite{Calefato:2018:IST}\end{tabular}} & \multicolumn{2}{c}{Senti4SD} & \multicolumn{2}{c}{SentiStrength-SE} & \multicolumn{2}{c}{DEVA} & \multicolumn{2}{c}{SentiCR(binary)}\\
				&\multicolumn{1}{c}{} & Coeff. & OR & Coeff. & OR & Coeff. & OR & Coeff. & OR & Coeff. & OR \\
				\hline 
				& (Intercept) & -1.60 & -- & -1.70 & -- & -1.74 & -- & -1.76 & -- & -1.74 & -- \\
				\hline
				\multirow{ 2}{*}{Affect} & Positive Sentiment & \textbf{-0.06} & 0.94 & \textbf{-0.06} & 0.94 & -0.03 & 0.97 & 0.00 & 1.01 & -- & -- \\
				& Negative Sentiment & \textbf{-0.21} & 0.81 & \textbf{-0.16} & 0.85 & -0.01 & 0.99 & 0.04 & 1.04 & -0.09 & 0.92 \\
				\hline
				\multirow{ 4}{*}{Time}& Weekend & \textbf{0.10} & 1.10 & \textbf{0.10} & 1.11 & \textbf{0.10} & 1.11 & \textbf{0.10} & 1.11 & \textbf{0.10} & 1.11 \\
				%GMTHour ('Morning' as default) & & & & & & & & & & & \\
				& GMTHour Afternoon & \textbf{0.07} & 1.07 & \textbf{0.06} & 1.07 & \textbf{0.07} & 1.07 & \textbf{0.06} & 1.07 & \textbf{0.07} & 1.07 \\
				& Evening & \textbf{0.15} & 1.16 & \textbf{0.14} & 1.15 & \textbf{0.14} & 1.15 & \textbf{0.1}4 & 1.15 & \textbf{0.15} & 1.16 \\
				& Night & \textbf{0.13} & 1.14 & \textbf{0.13} & 1.14 & \textbf{0.13} & 1.14 & \textbf{0.13} & 1.14 & \textbf{0.1}3 & 1.14 \\
				\hline
				\multirow{ 8}{*}{Presentation Quality}& Presence of Code Snippet & \textbf{0.71} & 2.04 & \textbf{0.73} & 2.08 & \textbf{0.72} & 2.05 & \textbf{0.72} & 2.05 & \textbf{0.72} & 2.06 \\
				%Low Uppercase Ratio {[}0. 0.1{[} & 0.12 & 1.27 & \textless 2e-16 & 0.00 & 1.00 & 0.00 & 1.00 & 0.00 & 1.00 & 0.00 & 1.00 \\
				& Low Uppercase Ratio & \textbf{0.12} & 1.27 & 0.00 & 1.00 & 0.01 & 1.00 & 0.00 & 1.00 & 0.00 & 1.00 \\
				%Body Length (‘Short’ as default. where ‘Short’ is in {[}0. 90{[}) & & & & & & & & & & & \\
				& Body Length: & & & & &  &  &  &  &  &  \\
				& \-\ \-\ Medium Body [90. 200[ & \textbf{-0.21} & 0.81 & \textbf{-0.20} & 0.82 & \textbf{-0.22} & 0.80 & \textbf{-0.23} & 0.79 & \textbf{-0.20} & 0.82 \\
				& \-\ \-\ Long Body [200. $+\infty$[ & \textbf{ -0.42} & 0.65 & \textbf{-0.45} & 0.64 & \textbf{-0.46} & 0.63 & \textbf{-0.49} & 0.61 & \textbf{-0.44} & 0.65 \\
				%Title Length (‘Short’ as default. where ‘Short’ is in {[}0. 6{[}) & & & & & & & & & & & \\
				& Title Length: & & & & &  &  &  &  &  &  \\
				& \-\ \-\ Medium Title [6. 10[ & 0.05  & 1.05 & 0.05 & 1.05 & \textbf{0.05} & 1.06 & \textbf{0.05} & 1.06 & \textbf{0.05} & 1.05 \\
				& \-\ \-\ Long Title [10. $+\infty$[ & \textbf{0.05} & 1.05 & 0.05 & 1.06 & \textbf{0.06} & 1.06 & \textbf{0.06} & 1.06 & \textbf{0.06} & 1.06 \\
				%Presence of Multiple Tags (‘Single tag’ as default) & -0.12 & 0.88 & 1.10E-08 & -0.13 & 0.88 & -0.14 & 0.87 & -0.14 & 0.87 & -0.14 & 0.87 \\
				& Presence of Multiple Tags & \textbf{-0.12} & 0.88 & \textbf{-0.13} & 0.88 & \textbf{-0.14} & 0.87 & \textbf{-0.14} & 0.87 & \textbf{-0.14} & 0.87 \\
				& Presence of URLs & -0.01 & 0.99 & -0.01 & 0.99 & -0.02 & 0.98 & -0.02 & 0.98 & -0.01 & 0.99 \\
				% Asker Reputation (‘New’ as default) & & & & & & & & & & & \\
				\hline
				\multirow{ 3}{*}{Reputation}& Low [10. 1K[ & \textbf{0.87} & 2.39 & \textbf{0.87} & 2.39 & \textbf{0.88} & 2.40 & \textbf{0.88} & 2.41 & \textbf{0.88} & 2.41 \\
				& Established [1K. 20K[ &\textbf{ 0.98} & 2.68 & \textbf{0.98} & 2.68 & \textbf{0.99} & 2.69 & \textbf{0.99} & 2.70 & \textbf{1.00} & 2.71 \\
				& Trusted [20K. $+\infty$[ & \textbf{1.17} & 3.23 & \textbf{1.17}& 3.21 & \textbf{1.18} & 3.25 & \textbf{1.18} & 3.26 & \textbf{1.19} & 3.30 \\
				\hline
				\multicolumn{12}{l}{\footnotesize{\textit{Boldface indicates statistical significance with} $p<.05$}}
			\end{tabular}
			
			\caption{Results of the replication of the study by ~\cite{Calefato:2018:IST} on success factors for Stack Overflow questions.} 
			\label{tab:ISTreplications}
		\end{table*}
\end{landscape}}

\begin{table}[tb]
	\begin{center}
		\caption{Distribution of polarity labels predicted by the SE-specific tools for the study by~\cite{Calefato:2018:IST}. The larger group per study is typeset in boldface.}
		\label{tab:labelDistributionIST}
		\begin{tabular}{l|lll}
			\textbf{Tool}    & \textbf{Negative} & \textbf{Neutral}   & \textbf{Positive}  \\
			\hline 
			\multicolumn{4}{c}{\textit{Original study by~\cite{Calefato:2018:IST}}}\\
			SentiStrength  & 11\% & 63\% & 26\% \\
			\hline
			\multicolumn{4}{c}{\textit{Our replications with SE-specific tools}}\\
			SentiStrength-SE & 17.00\%              &\textbf{51.07\% }              & 31.93\%               \\
			DEVA             & 20.27\%              & 40.05\%               & 39.68\%               \\
			SentiCR          & 39.33\%              & \multicolumn{2}{c}{\textbf{60.67\%} (non-negative)} \\
			Senti4SD         & 34.39\%             &\textbf{40.73\%   }            & 24.88\% \\
			\hline
		\end{tabular}
	\end{center}
\end{table}

In Table~\ref{tab:ISTreplications}, we report the results of the original study as well as the outcome of the replications with the four SE-specific sentiment analysis tools. For each predictor, we report the coefficient estimate, the odds ratio (OR), and indicate statistical significance of the correlation (p-value \textless .05) in bold. The sign of the coefficient estimate indicates the positive/negative association of the predictor with the success of a question. The odds ratio (OR) weighs the magnitude of this impact, with values close to 1 indicating a small impact. A value lower than 1 corresponds to a negative coefficients, and vice versa. Technically speaking, an $OR = x$ indicates that the odds of the positive outcome are $x$ times greater than the odds of the negative outcome. OR is an asymmetrical metric, with positive odds varying from 1.0 to infinity and decreasing OR bounded by 0. 
To further investigate the outcome of classification performed with the different tools, in Table~\ref{tab:labelDistributionIST}, we report the label distributions for the predictions issued by the SE-specific tools and provide comparison with those obtained with SentiStrength from the original study.

As for RQ1, we confirm the original findings related to \textit{Reputation}, \textit{Time}, and \textit{Presentation Quality}, for which we observe comparable coefficients and ORs in all settings. More in detail, we confirm that Reputation is the most influential factor for the success of questions, with \textit{Trusted} users having the highest probability of getting an accepted answers. As for Time, \textit{evening} and \textit{night} are confirmed as the most effective time zones. As far as Presentation Quality is concerned, the presence of code snippet remains the strongest predictor in all settings. The results from the replications also confirm the \textit{Body Length} to be negatively correlated with the success of questions. Conversely, we could not confirm the positive correlation between low uppercase ratio with the success of questions. 
As for the impact of the \textit{Affect} factor, we confirm the negative impact of both positive and negative sentiment on the success of questions when Senti4SD is used. Conversely, for SentiStrength-SE, DEVA, and SentiCR we do not find empirical support for this claim. As already observed in the replication of the study by Pletea and colleagues (see Tables~\ref{tab:resultsCommit} and~\ref{tab:resultsPullRequest} in Section~\ref{sec:resultsPletea}), the proportions of polarity labels vary depending on the tools.    
As such, we conclude that the choice of the sentiment analysis tools leads to partially contradictory results for this study. %(RQ1.1). 
The original findings about the impact of factors on the success of questions are mostly confirmed, with a couple of exceptions. Specifically, the sentiment from the original work are fully confirmed only when Senti4SD is used for performing sentiment analysis of questions, while the impact of writing in uppercase is not confirmed in any setting. %(RQ1.2).

\MyBox{\emph{\textbf{RQ1 Summary:}} The `off-the-shelf' use of SE-specific sentiment analysis tools may induce a threat to conclusion validity depending on the chosen level of granularity of the analysis. In both our replications, the tools lead to the same conclusions at a coarse level of granularity while differences are observed in terms of distribution of polarity labels in the two datasets.
	
	Our replications with SE-specific sentiment analysis tools confirm the findings from earlier studies at a coarse-grained level of analysis. 
	However, we find that the off-the-shelf use of sentiment analysis tools might lead to different results at a fine-level of granularity due to differences in the label distributions.}

\section{Agreement between SE-specific tools}
\label{sec:agreement}

To address RQ2 (\textit{To what extent do results from different SE-specific sentiment analysis tools agree with each other?}), we measure the agreement among the four SE-specific tools using the weighted kappa ($\kappa$) by~\cite{CohenJ1968}. The $\kappa$ metric is computed as the \textit{observed agreement}, i.e., the number of times the raters (either humans or classifiers) issue the same labels, corrected by the \textit{chance agreement}, that is the probability that the raters agree by chance. Consistently with previous research~\citep{Jongeling:2017,Novielli:2018}, we distinguish between \textit{mild disagreement}, that is the disagreement between negative/positive and neutral annotations, and \textit{severe disagreement}, that is the disagreement between positive and negative judgments. As such, in computing the weighted $\kappa$, we assigned a weight = 2 to severe disagreement and a weight = 1 to mild disagreement (see Table~\ref{tab:weights}). We follow the interpretation of $\kappa$ by~\cite{Viera2005UnderstandingIA}, suggesting that the agreement is less than chance if $\kappa \leq 0$, slight if $0.01 \leq \kappa \leq 0.20$, fair if $0.21 \leq \kappa \leq 0.40$, moderate if $0.41 \leq \kappa \leq 0.60$, substantial if $0.61 \leq \kappa \leq 0.80$, and almost perfect if $0.81 \leq \kappa \leq 1$.

\begin{table}[htb]
	\begin{center}
		\caption{Weighting scheme for computation of $\kappa$}
		\label{tab:weights}    
		\begin{tabular}{l|lll}
			& Negative & Neutral & Positive \\
			\hline
			Negative & 0        & 1       & 2        \\
			Neutral  & 1        & 0       & 1        \\
			Positive & 2        & 1       & 0       \\
			\hline
		\end{tabular}
	\end{center}
\end{table}

The $\kappa$ index represents the \textit{de facto} standard used in research to assess the inter-rater agreement. However, a debate is still open about the limitations of the $\kappa$ statistic due to the presence of bias in the case of skewed distributions of labels \citep{diEugenio}. The main problem with $\kappa$ applied to computational linguistics and text classification tasks is that it is possible to observe low $\kappa$ values even in presence of high agreement if the distribution of one label largely outnumbers the others. In fact, the minimum value of chance agreement occurs when the labels are equally distributed among all the possible categories (i.e., \textit{positive, negative}, and \textit{neutral} in our case). Conversely, the maximum value of chance agreement (i.e., 1) occurs when all the labels belong to a single category. Indeed, the problem of unbalanced polarity labels affects both our replication datasets, for which we observe a prevalence of neutral cases (see Tables~\ref{tab:resultsCommit},
\ref{tab:resultsPullRequest}, and \ref{tab:labelDistributionIST}). 

Given these limitations and for the sake of completeness, in Tables~\ref{tab:agreementPletea} and \ref{tab:agreementIST} we report both the weighted Cohen's $\kappa$ for the two datasets and the observed agreement, i.e., the the percentage of cases for which the tools in each pair (by row) issue the same prediction (perfect agreement). Furthermore, in line with previous research by~\cite{Novielli:2018}, we also report the percentage of cases for which severe and mild disagreement are observed. The only exception is represented by SentiCR that predicts two classes (i.e., negative vs. non-negative), thus making impossible to distinguish between severe and mild disagreement. 

For both datasets (see Tables~\ref{tab:agreementPletea} and \ref{tab:agreementIST}), the $\kappa$ values indicate an agreement between tools ranging from slight to moderate. We observe the highest values of $\kappa$ between SentiStrength-SE and DEVA is the highest (.45--.59). This is reasonable as they have been created by the same team and share the same underlying API and lexicon, which is an adaptation of the lexicon and the rules of SentiStrength. The second highest agreement (.33--.53) is observed between SentiStrength-SE and Senti4SD, which is again expected as Senti4SD also leverages the SentiStrength lexicon that significantly overlaps with the adapted version implemented by SentiStrength-SE. The lowest agreements involve SentiCR with both DEVA (.04--.18) and SentiStrength-SE (.08--.15). A possible explanation for this is that SentiCR does not leverage any sentiment dictionary, thus relying only on Bag-of-Words (BoW) as features for the supervised classification model trained on a gold standard of code reviews. Overall, Senti4SD is in the middle of the scale of observed $\kappa$ values, showing a moderate agreement with lexicon-based tools and a fair agreement with SentiCR, probably because it leverages both lexicon-based features and BoWs. 

These results are in line with findings from our previous benchmarking studies~\cite{Novielli:2018:MSR,Novielli:2020}. 
Furthermore, we observe an improvement from $\kappa = .25$, which is the highest agreement reported by~\citeauthor{Jongeling:2017} for general purpose tool on GitHub comments, to  $\kappa = .59$ and  $\kappa = .57$, which we observe for SentiStrength and DEVA, respectively on pull request and commit comments, thus confirming the previous findings about the positive impact of SE-specific tuning of sentiment analysis tools~\cite{Novielli:2018:MSR}. However, lower agreement is observed for longer texts such as GitHub discussions or Stack Overflow questions, for which we observe values of $\kappa=.47$ and $\kappa = .48$, respectively.

Looking at the \textit{disagreement} rates, we observe that the main cause for disagreement is between neutral and either positive or negative polarity (mild disagreement). The highest rates of severe disagreement are observed for both pull request and commit discussions (between 8--16\%, see Table~\ref{tab:agreementPletea}) as well as for Stack Overflow questions (between 7--17\%, see Table~\ref{tab:agreementIST}). For pull request and commit comments, the severe disagreement ranges between 2--6\%, thus providing a further indication of the higher difficulty of issuing an overall polarity label for longer texts. 

Finally, the $\kappa$ values are generally lower than the ones observed with fine-tuned SE-specific tools in a within-platform condition. In our previous study ~\cite{Novielli:2020}, we observed a drop in agreement between tools for the supervised classifiers (i.e., Senti4SD and SentiCR, in this case) when trained and tested in a cross-platform condition, that is, when using train and test sets from different platforms. In line with evidence provided in previous benchmark study, we observe the lowest agreement for pairs that include SentiCR. This might be explained by the approach implemented by SentiCR, which relies on features based on bag-of-words that hold a lower ability to generalize across-dataset thus causing overfitting to the training data of the platform-specific lexicon. This evidence confirms that shifts in lexical semantics also occur within the software development domain due to platform-specific jargon and communication style, and highlight that SE-specific classifiers are no silver bullet since problems arise if they are also used as `off-the-shelf' solutions.

\begin{table}[htb]
	\begin{center}
		\caption{Agreement between SE-specific tools for the dataset used in the study by~\cite{Pletea:2014}.}
		\label{tab:agreementPletea}
		\begin{tabular}{lcccc}
			&& \textbf{Perfect} &  \multicolumn{2}{c}{\textbf{Disagreement}} \\
			\textbf{Tool Pair}  & \textbf{$\kappa$}    & \textbf{Agreement}  & \textbf{Severe} & \textbf{Mild}  \\
			\hline
			\multicolumn{5}{c}{\textbf{Pull request discussions}}\\
			SentiStrength-SE vs. DEVA     & 0.45 & 67\%              & 10\%   & 23\% \\
			SentiStrength-SE vs. Senti4SD & 0.33 & 61\%              & 10\%   & 30\% \\
			DEVA vs. Senti4SD             & 0.29 & 58\%              & 16\%   & 26\% \\
			SentiStrength-SE vs. SentiCR  & 0.15 & 55\%              & \multicolumn{2}{c}{45\%}\\
			DEVA vs. SentiCR              & 0.25 & 61\%              & \multicolumn{2}{c}{39\%}\\
			SentiCR vs. Senti4SD          & 0.32 & 65\%              & \multicolumn{2}{c}{35\%}\\
			\hline
			\multicolumn{5}{c}{\textbf{Commit discussions}}\\
			SentiStrength-SE vs. DEVA     & 0.47 & 66\%              & 10\%   & 23\% \\
			SentiStrength-SE vs. Senti4SD & 0.45 & 66\%              & 8\%    & 26\% \\
			DEVA vs. Senti4SD             & 0.35 & 60\%              & 16\%   & 24\% \\
			SentiStrength-SE vs. SentiCR  & 0.15 & 63\%              & \multicolumn{2}{c}{36\%}\\
			DEVA vs. SentiCR              & 0.18 & 62\%              & \multicolumn{2}{c}{38\%}\\
			SentiCR vs. Senti4SD          & 0.29 & 69\%              & \multicolumn{2}{c}{31\%}\\
			\hline
			\multicolumn{5}{c}{\textbf{Pull request comments}}\\
			SentiStrength-SE vs. DEVA     & 0.59 & 84\%              & 2\%    & 14\% \\
			SentiStrength-SE vs. Senti4SD & 0.47 & 82\%              & 2\%    & 17\% \\
			DEVA vs. Senti4SD             & 0.36 & 75\%              & 3\%    & 22\% \\
			SentiStrength-SE vs. SentiCR  & 0.08 & 77\%              & \multicolumn{2}{c}{23\%}\\
			DEVA vs. SentiCR              & 0.05 & 74\%              & \multicolumn{2}{c}{26\%}\\
			SentiCR vs. Senti4SD          & 0.14 & 79\%              & \multicolumn{2}{c}{21\%}\\
			\hline
			\multicolumn{5}{c}{\textbf{Commit comments}}\\
			SentiStrength-SE vs. DEVA     & 0.57 & 76\%              & 4\%    & 20\% \\
			SentiStrength-SE vs. Senti4SD & 0.53 & 76\%              & 3\%    & 21\% \\
			DEVA vs. Senti4SD             & 0.39 & 66\%              & 6\%    & 28\% \\
			SentiStrength-SE vs. SentiCR  & 0.09 & 75\%              & \multicolumn{2}{c}{25\%}\\
			DEVA vs. SentiCR              & 0.04 & 69\%              & \multicolumn{2}{c}{31\%}\\
			SentiCR vs. Senti4SD          & 0.16 & 77\%              & \multicolumn{2}{c}{23\%}\\
			\hline
		\end{tabular}
	\end{center}
\end{table}

\begin{table}[htb]
	\begin{center}
		\caption{Agreement between SE-specific tools for the dataset used in the study by~\cite{Calefato:2018:IST}.}
		\label{tab:agreementIST}
		\begin{tabular}{lcccc}
			&& \textbf{Perfect} &  \multicolumn{2}{c}{\textbf{Disagreement}} \\
			\textbf{Tool Pair}  & \textbf{$\kappa$}    & \textbf{Agreement}  & \textbf{Severe} & \textbf{Mild}  \\
			
			\hline
			SentiStrength-SE vs. DEVA     & 0.48  &   68\%             & 7\%   &25\% \\
			SentiStrength-SE vs. Senti4SD &  0.28 &    53\%            & 11\%  & 36\% \\
			DEVA vs. Senti4SD             &  0.21 &    50\%            & 17\%  & 34\% \\
			SentiStrength-SE vs. SentiCR   & 0.10  &    61\%              & \multicolumn{2}{c}{39\%}\\
			DEVA vs. SentiCR               &  0.14 &        63\%           & \multicolumn{2}{c}{37\%}\\
			SentiCR vs. Senti4SD           &  0.23 &         64\%            & \multicolumn{2}{c}{36\%}\\
			\hline
		\end{tabular}
	\end{center}
\end{table} 

\MyBox{\emph{\textbf{RQ2 Summary:}} Agreement between SE-specific tools ranges from slight to moderate. The highest agreement is observed for the two tools sharing the same lexicon and developed upon the same API (i.e., SentiStrength-SE and DEVA). More cases of severe disagreements are observed in GitHub discussions and Stack Overflow questions than in GitHub comments, thus suggesting that issuing an overall polarity label is more difficult for longer texts.}

\section{Agreement of SE-specific tools with manual annotation}
\label{sec:manual}

To address RQ3 (\textit{To what extent do different SE-specific sentiment analysis tools agree with the emotions of software developers?}), we manually labeled a subset of 600 documents randomly selected from the GitHub and Stack Overflow datasets: 200 pull request and 200 commit comments, equally distributed between the `security' and `no security' groups from the dataset by \cite{Pletea:2014}, plus 200 questions from the dataset by \cite{Calefato:2018:IST}

The labeling study was performed by the first two authors (\textit{raters} hereinafter), following previously published annotation guidelines for sentiment polarity by~\cite{Calefato:2018}. Each document in the subsets was individually annotated by the two raters. 
For each document, the raters indicated whether positive or negative sentiment was conveyed. It was also possible to indicate both positive and negative polarity, to represent cases of \textit{mixed} polarity. To indicate neutral sentiment, the raters were required to annotate absence of both positive and negative polarity. 
All the cases of disagreement were resolved through discussion, leading to the assignment of a gold label to each document. Two comments from the GitHub subset were discarded because the raters could not reach an agreement on the polarity label. Conversely, the raters successfully addressed all the disagreements in the Stack Overflow subset. 
The resulting label distributions for the two subsets are comparable, with the majority of cases labeled as neutral (see Table~\ref{tab:labelDistributionManual}). The GitHub subset contains a slightly higher proportion of positive cases compared to the Stack Overflow one. Vice versa, negative sentiment occurs more often in Stack Overflow. A larger proportion of mixed cases appear in the Stack Overflow subset (7\%) than in GitHub  (2\%). A possible explanation for this, as also emerged during the disagreement resolution discussions, is that longer texts are more likely to convey both positive and negative sentences. In fact, the average length of the documents in the GitHub sample is 134 characters whereas it is 430 in Stack Overflow.  

We measure the agreement between the raters, as well as between manual gold labels and the tool predictions, using the same metrics and approach adopted for assessing the agreement between tools (see Section~\ref{sec:agreement}). 
The agreement metrics in Table~\ref{tab:agreementManual} show a moderate agreement between the two raters for both datasets. Similarly to what observed for the agreement between tools, $\kappa$ values indicate better agreement for shorter documents. This holds true for both inter-rater agreement in the manual labeling study and for the tool-based vs. manual labels. In fact, we observe better agreement for the GitHub commit and pull request comments (average length of the comments in the labeled set is 134 characters).  Conversely, a lower agreement is observed for the Stack Overflow questions (average length of 430 characters) for which we also observe a larger proportion of severe disagreement cases and mixed cases (see Table~\ref{tab:labelDistributionManual}).

\begin{table}[bt]
	\begin{center}
		\caption{Distribution of polarity labels resulting from the manual annotation.}
		\label{tab:labelDistributionManual}
		\begin{tabular}{l|rr}
			& \multicolumn{2}{c}{\textbf{Dataset}}\\
			\hline
			& \textit{Pletea et al.} & \textit{Calefato et al.} \\
			
			Unit of annotation & GitHub comments                 & Stack Overflow questions           \\
			\hline
			\#Documents & 400                    & 200                    \\
			Negative    & 47 (12\%)                  & 35 (18\%)                     \\
			Neutral    & 290 (73\%)                   & 132 (66\%)                     \\
			Positive     & 52 (13\%)                  & 20 (10\%)                      \\
			Mixed        & 9 (2\%)                   & 13 (7\%)                     \\
			Discarded        & 2 (.05\%)                 & ---                    \\ 
			\hline
		\end{tabular}
	\end{center}
\end{table}

%\input{tables/agreement_with_manual_labelling.tex}

% \usepackage{graphicx}
\begin{table}[hbt]
	\begin{center}
		\caption{Agreement between raters in the labeling study and between tools and manual labels.}
		\label{tab:agreementManual}
		\resizebox{\textwidth}{!}{%
			\begin{tabular}{lrrrr}
				& \multicolumn{1}{l}{\textbf{Weighted}} & \multicolumn{1}{l}{\textbf{Observed}} & \multicolumn{2}{l}{\textbf{Disagreement}} \\
				\textbf{Pair} & \multicolumn{1}{l}{\textbf{k}} & \multicolumn{1}{l}{\textbf{Agreement}} & \multicolumn{1}{l}{\textbf{Severe}} & \multicolumn{1}{l}{\textbf{Mild}} \\ \hline
				\multicolumn{5}{c}{\textit{Github comments}} \\ \hline
				Rater 1 vs. Rater 2 (manual labeling) & 0.54 & 80\% & 3\% & 17\% \\ \hline
				SentiStrength-SE vs. manual labels & 0.50 & 77\% & 8\% & 21\% \\
				Senti4SD vs. manual labels & 0.44 & 72\% & 3\% & 24\% \\
				DEVA vs. manual labels & 0.35 & 69\% & 6\% & 25\% \\
				SentiCR vs. manual labels & 0.16 & 81\% & \multicolumn{2}{c}{19\%} \\ \hline
				\multicolumn{5}{c}{\textit{Stack Overflow questions}} \\ \hline
				Rater 1 vs. Rater 2 (manual labeling) & 0.45 & 69\% & 3\% & 27\% \\ \hline
				SentiStrength-SE vs. manual labels & 0.29 & 57\% & 5\% & 37\% \\
				Senti4SD vs. manual labels & 0.29 & 59\% & 7\% & 33\% \\
				DEVA vs. manual labels & 0.19 & 51\% & 11\% & 37\% \\
				SentiCR vs. manual labels & 0.13 & 68\% & \multicolumn{2}{c}{32\%} \\ \hline
				& \multicolumn{1}{l}{} & \multicolumn{1}{l}{} & \multicolumn{1}{l}{} & \multicolumn{1}{l}{} \\
				& \multicolumn{1}{l}{} & \multicolumn{1}{l}{} & \multicolumn{1}{l}{} & \multicolumn{1}{l}{}
			\end{tabular}%
		}
	\end{center}
\end{table}

\subsection{Follow-up analysis on majority voting}
We run a follow-up analysis to investigate if the agreement with the manual gold label improves if we focus on the comments for which the tools agree with each other. To this aim, we replicate the approach followed by~\citeauthor{Jongeling:2017} and select the subset of GitHub comments and Stack Overflow questions for which Senti4SD, SentiStrength-SE, and DEVA agree on the polarity classification. We cannot include SentiCR in this analysis as we are not able to disambiguate between positive and neutral cases for which the tool issues the `non-negative' macro-label. As a further investigation, we also repeat the same assessment by considering only the two tools with higher agreement with manual labels, namely SentiStrength-SE and Senti4SD (see Table~\ref{tab:agreementManual}). Of course, given the observed agreement between tools, this analysis focuses on a reduced number of text items. We compute the agreement of tools with the manual labels on 262 comments from the GitHub sample (66\%) and 73 questions from  Stack Overflow (36\%). %This difference in proportion of texts for which the three tools agree reinforces the previous evidence suggesting that agreement between tools and with manual labels is easier to achieve for shorter texts. 
For two tools, a larger proportion of texts is included in the analysis, i.e., 302 GitHub comments (75\%) and 111 Stack Overflow questions (56\%).

We summarize the results of this follow-up study in Table~\ref{tab:threeTolsWithManual}.  As already reported by \citeauthor{Jongeling:2017}, we observe that focusing on the comments where the tools agree improves the agreement with the manual labeling, both in terms of weighted $\kappa$ and in terms of disagreement rates. Agreement with manual labels become substantial, raising from a maximum of .50 in the full dataset (SentiStrength-SE, see Table~\ref{tab:agreementManual}) to .69 for GitHub and from .29 (SentiStrength-SE and Senti4SD, see Table~\ref{tab:agreementManual}) to .60 for Stack Overflow. As for disagreement, the proportion of severe disagreement rates drops to 2\% and 3\% for GitHub and Stack Overflow, respectively, which is comparable to the disagreement between human raters (see Table~\ref{tab:agreementManual}). 
More recently, similar findings were presented by~\cite{Zhang:ICSME2020}. In their study leveraging deep learning for sentiment analysis, they provided evidence on how composition of different classifiers may boost performance.
In line with previous findings, this evidence suggests that in absence of a gold standard for retraining, implementing a voting system between off-the-shelf tools might be a way to increase classification performance. However, a trade-off exists between the accuracy and the agreement with manual labeling. This approach would not enable to classify the polarity of sentiment for those cases for which tools do not reach an agreement, thus leaving the sentiment for such documents undetermined.  

\begin{table}[tb]
	\caption{Agreement of group of tools with manual labeling.}
	\label{tab:threeTolsWithManual}
	\resizebox{\textwidth}{!}{%
		\begin{tabular}{llllll}
			\multirow{2}{*}{\textbf{Tools}} & \multicolumn{1}{c}{\textbf{\#agreement}} & \multicolumn{1}{c}{\textbf{Weighted}} & \multicolumn{1}{c}{\textbf{Observed}} & \multicolumn{2}{c}{\textbf{Disagreement}} \\
			& \multicolumn{1}{c}{\textbf{cases (\%)}} & \multicolumn{1}{c}{\textbf{k}} & \multicolumn{1}{c}{\textbf{Agreement}} & \multicolumn{1}{c}{\textbf{Severe}} & \multicolumn{1}{c}{\textbf{Mild}} \\ \hline
			\multicolumn{6}{c}{\textit{GitHub}} \\
			Three tools & 262 (66\%) & .69 & 87\% & 2\% & 11\% \\
			Two tools & 302 (75\%) & .60 & 85\% & 2\% & 12\% \\
			\multicolumn{6}{c}{\textit{Stack Overflow}} \\
			Three tools & 73 (36\%) & .60 & 78\% & 3\% & 19\% \\
			Two tools & 73 (36\%) & .60 & 78\% & 3\% & 19\% \\
			\hline
		\end{tabular}%
	}
\end{table}

\subsection{Error analysis}
To better understand the difficulties inherent to sentiment detection in our dataset, we performed a qualitative analisys by manually examining the cases for which the gold label did not match the majority voting label based on tool predictions. In order to include SentiCR in this analysis, we mapped positive and neutral labels to `non-negative'. Furthermore, we excluded from this analysis the text documents for which we could not assign a polarity class based on majority voting between tools (31 for GitHub and 34 for Stack Overflow). The resulting collection counted 67 text documents (referred to as the \textit{disagreement set} hereinafter), of which 36 are GitHub comments and 30 are Stack Overflow questions. The two authors who performed the manual labeling also inspected the content of the documents in the disagreement set to identify the cause of error. Specifically, one rater assigned the error categories, while the other one reviewed  them. The two raters labeled the disagreement set by using the error categories defined in the benchmarking study on SE-specific sentiment analysis tools by~\cite{Novielli:2018:MSR}. Therefore, each document in the disagreement set was annotated with possible causes of error. In Table~\ref{tab:errorDistribution}, we report the error category distribution resulting from our analysis, ordered by frequency of observation. The overall number of documents belonging to each class is reported as well, with a breakdown by dataset. For each error category, we also indicate the percentage of items belonging to it. In 7 cases multiple error categories were selected, indicating multiple causes for misclassification. For this reasons the percentages do not sum up to 100\%. 
In 9\% of cases (10 Stack Overflow questions), the raters agreed that there was no error in the classification as the disagreement actually originated from the document conveying both positive and negative polarity (i.e., \textit{mixed} cases in Table~\ref{tab:errorDistribution}), and, therefore, they are not included in the error category count. 

\begin{table}[tb]
	\caption{Distribution of error categories in the Github and Stack Overflow (SO) datasets.}
	\label{tab:errorDistribution}
	\begin{tabular}{l|ccc}
		\textbf{Error category}     & \textbf{Github} & \textbf{SO} &  \textbf{overall (\%)} \\
		\hline 
		General error               & 12               & 17           & 29 (43\%)             \\
		Implicit sentiment polarity & 13              & 8            & 21 (31\%)             \\
		Figurative language         & 5               & 1            & 6 (9\%)              \\
		Politeness                  & 1               & 3           & 4 (6\%)             \\
		Pragmatics                  & 3               & 1            & 4 (6\%)             \\
		Subjectivity in annotation  & 2               & -            & 2 (3\%)              \\
		Polar facts                 & 1               & -           & 1 (1\%)          \\
		
		\hline 
		%Mixed polarity (no error)           & 9               & 13           &               \\
		Overall labeled documents           & 35              & 30           &                  \\
		Overall labels assigned              & 37              & 30           &                  \\
		Documents with multiple error categories     & 2               & -            &           \\
		\hline
	\end{tabular}
\end{table}

The most frequent error category is \textit{general error}, that is the inability to identify lexical cues of sentiment, as in the following GitHub comment:
\begin{adjustwidth}{0.7cm}{}
	``\textit{The alternative of course is to fixup the build system to split swscale and libav*, which wouldn't be a terrible idea as there are several places currently where the removal of ffmpeg would mean the removal of unrelated features (screenshots, as an example).}''
\end{adjustwidth}
that is erroneously classified as non-negative by the majority of tools, probably due to their inability to deal with the negation (\textit{wouldn't} of the negative word \textit{terrible}). Other causes for general errors are related to the wrong preprocessing of the raw text (i.e., code or URL not removed properly before launching the tool-based classification), or to the fact that the classifiers may overlook lexical features  because either they are not included in their sentiment lexicons 
%of SentiStregth-SE and DEVA 
or they are not modeled as relevant due to the low frequency in the training sets of the supervised classifiers.

In 31\% of cases the tools fail because of the presence of \textit{implicit sentiment polarity} in text, as in as in the following examples: 

\begin{adjustwidth}{0.7cm}{}
	``\textit{Any way to report as spam on comments like these}'' 
\end{adjustwidth}

\begin{adjustwidth}{0.7cm}{}
	\textit{``Recently there were quite some changes which have not made their way to the wiki yet. Someone should update it...''} 
\end{adjustwidth}

\begin{adjustwidth}{0.7cm}{} 
	\textit{``It is required to return HashWithIndifferentAccess instead of just plain regular Hash??? I mean what value does it bring in to the code???''}
\end{adjustwidth}

These cases received a \textit{negative} gold label because human raters could identify the negative attitude of the authors' comments  towards their interlocutor, even in the absence of explicit emotion lexicon. This evidence is in line with previous findings from our study on anger in collaborative software development, showing how detecting negative sentiments towards peers is more difficult than detecting anger towards self or objects~\citep{Gachechiladze:2017}. In fact, we observed that a hostile attitude towards the interlocutor is often phrased using implicit, indirect lexicon, probably to avoid too aggressive verbal behavior.

In 9\% of misclassified texts, we observe the use of \textit{figurative language} such as the humor, irony, sarcasm, or metaphors. For example:

\begin{adjustwidth}{0.7cm}{}
	``\textit{@username  oh, duh! I'll just go over to the corner and hang my head for awhile for forgetting about that. I suppose it may be still worth adding a method to do the comparison, as a convenience and `nudge in the hey use this direction' for plugin a}.''
\end{adjustwidth} 

\begin{adjustwidth}{0.7cm}{}
	``\textit{You are getting rusty in your old age...}.''
\end{adjustwidth} 
Such cases are extremely challenging to classify for sentiment analysis tools because much figurative language is based on conventions, such as idiomatic expressions and proverbs. Indeed dealing with figurative language, in general, and irony and sarcasm in particular, is still an open challenge for sentiment analysis research~\cite{bosco_patti, Weitzel2016}.

A few cases are misclassified because of the presence of ambiguous lexical cues of \textit{politeness} such as `\textit{thank you},' which might be either interpreted as a way to convey politeness or actual emotions (e.g., showing gratitude or appreciation)~\citep{Novielli:2015}. Finally, we observe a few cases for which misclassification is due to the inability of tools to deal with \textit{pragmatics}. It is the case, for example, of suggestions formulated as questions, as in ``\textit{Would IE user agent sniffing be a bad idea?},'' where the presence of negative lexicon (i.e., \textit{bad}) causes misclassification of this neutral text as negative. This is in line with previous research showing how developers use questions to express emotional lexicon in order to induce critical reflection in the interlocutor, to express criticism but also actual emotions, such as anger or surprise, as well as to communicate perplexity or disagreement with respect to the implemented solution~\citep{Ebert:2018}. Being able to identify the actual communicative intention of a questions is not a trivial task. It requires being able to understand information about context and pragmatics that cannot be successfully processed by state-of-the-art sentiment analysis tools, which rather rely on lexical semantics of documents.

\MyBox{\emph{\textbf{RQ3 Summary:}} The agreement between manual and tool annotation is higher for shorter documents (GitHub comments). The main causes of error are the presence of lexical cues of sentiment that are either wrongly processed or overlooked by the tools, the use of neutral lexicon to implicitly convey non-neutral polarity, and the use of figurative language.}

\section{Discussion}
\label{sec:discussion}

In the following, we summarize the key insights from our replications in the form of actionable guidelines to inform future research on sentiment analysis in software engineering.

\textit{Sentiment analysis tools should be retrained, if possible, rather than used off the shelf.} 
We observe that,  when used off-the-shelf as in our replications, SE-specific sentiment analysis tools may lead to contradictory results if different levels of unit of analysis are considered.  
When replicating the study by Pletea et al., we could confirm the original findings that GitHub security comments and discussions convey more negative sentiment than non-security ones, regardless of the tool used. Similarly, when replicating the study on Stack Overflow by Calefato et al., we could confirm the impact on question success of users' reputation and the presentation quality of questions. However, in both cases, we observe a different distribution of polarity labels that may lead to different conclusions at a finer-grained level of analysis. 
As such, albeit specifically tuned for the software engineering domain,  we argue that using sentiment analysis tools off-the-shelf poses a potential threat to conclusion validity, which might have an impact on the findings of previously published studies in the field. This evidence is also supported by the findings reported in a benchmarking study by \cite{Novielli:2020}, who report a drop in performance in cross-platform settings, i.e., when SE-specific sentiment analysis tools are used off-the-shelf in the absence of a gold standard for retraining. 

\textit{Perform a preliminary sanity check to select the appropriate tool in line with the research goals.} One of the assumptions underlying the choice of a sentiment analysis tool is that we share the original goal and the same conceptualization of affect with the authors of the tool. This is not necessarily true, regardless of the fine-tuning for the SE domain, as different gold standard might be inspired by different theoretical models of affect, aiming at modeling different affective states, such as emotions, interpersonal stances, attitudes, or moods. 
The disagreement between tools observed in our replications shows how SE-specific tuning of sentiment analysis tools does not necessarily represent \textit{per se} a silver bullet for improving the accuracy of sentiment analysis tools in software engineering studies. Therefore, a possible explanation for the low agreement observed is that the benchmarked tools have been originally validated and tuned on gold standards that include manual annotation following different guidelines. 
As already pointed out by previous research, sentiment annotation is a subjective task, thus even humans might disagree with each-others~\citep{Imtiaz:2018} if model-driven annotation is not adopted~\citep{Novielli:2018}. 
Moreover, \cite{Islam:Zibran:benchmark} showed how  tools exhibit their best performance on the dataset they were originally tested at the time of their release, whereas a drop in performance is observed when they are assessed on a different dataset. Furthermore, they report that the accuracy of the tools largely vary across different datasets in line with what was observed by~\cite{Lin:2018} and further confirmed by our cross-platform benchmarking study~\citep{Novielli:2020}. 

In our previous research, we already argued about the importance of grounding sentiment analysis research on theoretical models of affect~\citep{Novielli:2018}. As such, a sanity check is always recommended to assess the suitability of existing tools with respect to the specific research goals \citep{Novielli:2018:MSR, Novielli:IEEESW:2020}.
For example, in some cases the mining of opinions might be the target rather than the recognition of actual emotions, such as joy or sadness. \cite{Lin2019} reported poor performance of classifiers designed to detect developers' \textit{emotions} (e.g.,  \textit{love}, \textit{joy}, \textit{fear}) when applied to the different task of detecting developers' \textit{opinions} about software libraries. 

\textit{If retraining is not possible, consider an ensemble of multiple tools to improve performance}.
As for any classifier, the retraining of sentiment analysis tools on new datasets can largely improve their performance. However, not all the solutions support retraining, as in the case of the lexicon-based tools DEVA and SentiStrength-SE. 
Based on the results of our analysis on the agreement of tools (see Section~\ref{sec:agreement}) and in line with previous evidence \citep{Jongeling:2017,Zhang:ICSME2020}, we suggest implementing an ensemble of tools with a majority voting system as a possible way to increase the agreement with manual labels when the retraining of the selected solution is not an option. 

\textit{Beware of the effects of different units of analysis on tool performance.} The results of the agreement study suggest that the choice of the unit of analysis may represents a potential threat to construct validity. We observe better agreement for shorter documents both between tools (see Tables~\ref{tab:agreementPletea} and \ref{tab:agreementIST}) as well as between tools and manual annotation (see Table~\ref{tab:agreementManual}). Specifically, we found the best  $\kappa$ values of agreement between tools for GitHub comments (shorter unit of analysis), immediately followed by Stack Overflow posts, and the worst values for GitHub discussions (longer unit of analysis). This evidence suggests that sentiment analysis perform better on short text, whereas longer documents are more problematic as they might actually convey both positive and negative emotions--as confirmed by the higher percentage of manually-labeled mixed cases for the Stack Overflow dataset (7\%) compared to GitHub comments (2\%) (see Table~\ref{tab:labelDistributionManual}).  

Furthermore, we observe that each tool produces different distributions of positive, negative, and neutral labels for the dataset by Pletea et al., who considered as units of analysis individual comments vs. entire discussion. This confirms the evidence by \cite{Jongeling:2017} that care is needed when choosing the unit of analysis as it may affect the resulting polarity label distribution.
Therefore, other than fine-tuning sentiment analysis tools to address the challenges specific of the software engineering domain \citep{Novielli:2018} and the data source~\citep{Lin:2018}, tools might need further \textit{ad hoc} tuning if different lengths of documents or interaction threads are considered. For example, for the two supervised tools considered in this study (Senti4SD and SentiCR), this tuning could be done by retraining the classification models using a training set with gold labels assigned at a different granularity, e.g., at the level of individual sentences rather than at the level of the entire document. If retraining is not feasible due to the absence of a gold standard, we recommend explicit modeling of mixed cases by using tools that are designed to denote the presence/absence of positive and negative sentiment, such as SentiStrength-SE.

Finally, the dependency of the label distribution on the tools used and on the unit of analysis suggest further implications on the replicability of findings. In spite of the different polarity distribution labels as issued by the four tools inlcuded in our replications, the claims of Pletea et al. are more stable regardless of the tool used since they do not explicitly refer to this distribution but to percentage over groups (i.e., security vs. non-security) as opposed to negative vs positive. These findings are also confirmed when discussions rather than comments are used as a unit of analyis. However, different scenarios and research goals might suffer  seriously from the impact of the choice of the tools and of the unit of analysis. It is the case, for example, of identification of negative comments denoting a hostile attitude in the scope of community moderation activities, where optimizing by precision of the negative class on a single comment is crucial to avoid unnecessary ban of users.

\section{Threats to Validity}
\label{sec:threats}

Since we strictly replicated the two original studies, we also inherited some of the threats to validity reported in the original papers, e.g., the datasets under consideration are not representative for GitHub and Stack Overflow as a whole. Still, the limitations to the validity of each study are shared across both the original studies and the replications, and, therefore, unlikely to have influenced the findings reported here.

As for the annotation of the 600 document gold standard, threats to construct validity are mitigated by the fact that the labeling is performed by two independent raters, with discussion-based resolution of disagreements. 

When running replications, one of the inherent risk is to introduce unintended minor changes in the settings used in the original study, thus observing major differences in the results that are due to confounding factors. To mitigate this risk, we decided to perform our own studies, to avoid weakening the conclusion validity of our study. 
In particular, we performed two dependent replications in which the original design of the studies was preserved. The only change we introduced consists in the choice of the sentiment analysis tool, in line with our research goal of investigating the impact of the choice SE-specific tools on the conclusion validity.

%\section{Related work}
%\label{sec:related} 
%\input{sections/related.tex}

\section{Conclusion}
\label{sec:conclusion}

In this paper, we reported the results of an extended replication aimed at assessing to what extent SE-specific sentiment analysis tools mitigate  the  threats to conclusion validity highlighted by previous research. We found that, despite being tuned to address the challenges specific to the SE domain, the use of different sentiment analysis tools might lead to contradictory results when used off-the-shelf. This is especially true at a fine-grained level of analysis, as we found a moderate agreement between tools as well as differences in the distribution of polarity labels assigned by the different tools in the text items of our dataset.

Our results suggest that SE-specific fine-tuning of sentiment analysis tools to the software engineering domain might not be enough to improve accuracy. Conversely, platform-specific tuning or retraining might be needed to adjust the model performance to the shifts in lexical semantics due to different platform jargon or conventions. Further fine-tuning or retraining of tools might be required with respect to different lengths of documents as we found that longer texts might convey mixed sentiment polarity. Finally, a sanity check of the tool performance against manual labeling should always be performed as classifiers might be built according to different operationalization of emotions, which do not necessarily match the intended research goals. Dealing with figurative language and implicit sentiment polarity in texts still represent open challenges for sentiment classifiers that should be address by future work in this field. 

\section*{Acknowledgment}

We would like to thank Marika Redavid for her contribution in the early stage of this project.  The computational work has been executed on the IT resources made available by two projects, ReCaS and PRISMA, funded by MIUR under the program PON R\&C 2007–2013.

\bibliographystyle{spbasic}
\balance
\bibliography{references}

\end{document}